\documentclass[pre,twocolumn,superscriptaddress,showpacs]{revtex4}

\usepackage{graphicx} 
\usepackage{amsmath}  

\begin{document}    
 
\title{Force-Extension Relations for Polymers with Sliding Links}  
\author{Ralf Metzler} 
\email{metz@nordita.dk} 
\affiliation{Department of Physics, Massachusetts 
Institute of Technology, Cambridge, Massachusetts 02139} 
\affiliation{NORDITA, Blegdamsvej 17, DK-2100 Copenhagen {\O}, Denmark} 
\author{Yacov Kantor} 
\email{kantor@post.tau.ac.il} 
\affiliation{School for Physics and Astronomy, Tel Aviv 
University, Tel Aviv 69978, Israel} 
\affiliation{Department of Physics, Massachusetts 
Institute of Technology, Cambridge, Massachusetts 02139} 
\author{Mehran Kardar} 
\affiliation{Department of Physics, Massachusetts 
Institute of Technology, Cambridge, 
Massachusetts 02139} 
 
\date{\today}  
 
\begin{abstract}  
Topological entanglements in polymers are mimicked by sliding rings
({\em slip-links}) which enforce pair contacts between monomers.
We study the force-extension curve for linear polymers in which slip-links
create additional loops of variable size. 
For a single loop in a phantom chain, we obtain exact expressions 
for the average end-to-end separation: 
The linear response to a small force is related to the properties of the unstressed chain,
while for a large force the polymer backbone can be treated as a sequence of 
Pincus--de Gennes blobs, the constraint effecting only a single blob.
Generalizing this picture, scaling arguments are used to include self-avoiding effects. 
\end{abstract} 
 
 
\maketitle 
 
Entanglements play an important role in the behavior of macromolecules. 
For instance, mechanical links (e.g., in catenanes) and knots naturally appear in 
long polymers \cite{catenanes}. 
In biological systems, specific proteins act upon topological states: 
the degree of entanglement of chromosomes during cell division \cite{alberts}, 
or knotted states in bacterial DNA which may arise during random ring closure \cite{ringclosure}, 
can be modified by topological enzymes \cite{enzyme}. 
Synthetic RNA trefoil knots have been used to prove the existence of a similar, 
previously unknown, topology-changing enzyme \cite{wang}. 
Tight molecular knots have even been found deep inside the native state 
of proteins \cite{taylor}. 
Experimental advances now make it possible to manipulate single molecules 
by optical tweezers.
Thus, tight knots could be tied into single actin filaments or DNA strands \cite{arai}. 
Mechanical properties, and forces in the pN range relevant to biopolymeric processes,
can be measured by atomic force microscopy, or more direct 
micromechanical methods \cite{mechsing}. 
It is therefore possible to record the force-extension (FE) curve of single 
polymers with a fixed topology, from which valuable information about the 
properties of a molecule can be obtained and compared to theoretical predictions. 
 
\begin{figure} 
\includegraphics[width=4cm]{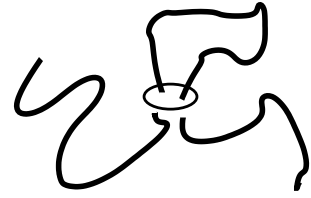} 
\caption{\label{link_def} Polymer threaded through a slip-link (ring) 
forms a loop and two dangling ends. It is not allowed to withdraw from 
the slip-link.} 
\end{figure}

While there has been extensive progress in the statistical mechanics of polymers
in the last decades \cite{degennesSC,edwards},
the analysis of topological constraints is hampered by the difficulty of treating 
the resulting division of phase space into accessible and inaccessible regions. 
Since the mathematical methods of knot detection 
using topological invariants \cite{mathdet} cannot be conveniently 
incorporated into a statistical-mechanical formulation, 
one may try to use geometrical constrictions to mimic knots: 
Consider a polymer threaded through a small ring as depicted 
in Fig.~\ref{link_def}, and not allowed to withdraw from it, although the 
ring may freely slide along the polymer and the loop size can change. 
Constrictions of this type (called slip-links (SLs)) were introduced  
by  Ball, Doi, Edwards and others \cite{slip}  to investigate the   
elasticity of rubber, where they were used to represent entanglements 
between different polymers. Recently, a detailed study of the size 
distribution of loops in SL structures was  performed \cite{metz}. In this work 
we consider FE curves of such polymers, with or without self-avoidance. 
We show that the knowledge of the statistics in the absence of the force, 
combined with the Pincus--de Gennes \cite{blob} 
blob picture, suffices to understand many features of the FE curves.

%%%%%%%%%%%%%%%%%%%%%%%%%%%%%%%%%%%%%%%%%%%%%%%%%%%%%%%%%%%%%%%%%%%%% 
%%%%%%%%%%%%%%%%%%%%%%%%%%%%%%%%%%%%%%%%%%%%%%%%%%%%%%%%%%%%%%%%%%%%% 
 
We first discuss the statistical effects of the slip-link 
for non-self-avoiding chains (phantom polymers). 
Consider an $N$-step chain with mean-squared step  
size $a^2$ in $d$-dimensional space.
In the $N\gg1$ limit, the probability density function
for the end-to-end distance (EED) {\bf r} is 
\begin{equation}\label{pN} 
p_N({\bf r})=\left(\frac{d}{2\pi Na^2}\right)^{d/2}\exp\left(-\frac{dr^2}{2Na^2}\right). 
\end{equation}  
This expression also describes an $N$-step random walk  
on a (hyper)cubic lattice with lattice constant $a$.
The number of closed $N$-step loops on such a lattice is $(2d)^N[d/(2\pi N)]^{d/2}$.  
Now consider the case where a SL  forces a closed loop of size $n$. 
Since the loop can be located on any of $N-n$ positions on the $N$-step chain, 
the number of possible configurations of the combined system of the polymer 
and the SL is $(N-n)(2d)^{N-n}(2d)^n[d/(2\pi n)]^{d/2}$. 
Thus, for $n\gg 1$ and $(N-n)\gg 1$, the probability for a given $n$ is 
\begin{equation}\label{gN} 
g_N(n)={\cal N}(N-n)n^{-d/2}\ , 
\end{equation} 
where ${\cal N}$ is a normalization factor.
For $d>2$, ${\cal N}$ depends on the short-distance cut-off, i.e., 
on the microscopic details of the walk \cite{d2}.
The presence of the SL also modifies the probability density of the EED, $p({\bf r})$,
by reducing the length of the backbone to $(N-n)$.
 
When the end-points of a polymer are stretched by a force ${\bf f}$, 
its properties can be derived from the partition function 
\begin{equation} 
\label{Z} 
{\cal Z}(f)=\int d^d{\bf r}\, p({\bf r})\,e^{{\bf f}\cdot{\bf r}/T}, 
\end{equation} 
where $f=|{\bf f}|$, and the temperature $T$ is in energy units,  
i.e., $k_B=1$. 
In the presence of the force, the vector ${\bf r}$ is on average parallel 
to ${\bf f}$, and from  Eq.~(\ref{Z}), its mean projection along ${\bf f}$ is
given by 
\begin{equation}\label{rfZ} 
\langle r\rangle_f=T\frac{\partial\ln{\cal Z}}{\partial f}. 
\end{equation} 
In particular, for $p({\bf r})=p_N({\bf r})$ in Eq.~(\ref{pN}), the partition function  
is ${\cal Z}=\exp[Na^2f^2/(2dT^2)]$, while the mean EED is a linear function 
of $f$, 
$\langle r\rangle_f=fNa^2/(dT)$, for any arbitrary value of the force. 
Since the mean-squared EED $R^2$ of an unstrained 
(zero-force) phantom chain is $Na^2$, the FE relation can be re-written as 
\begin{equation}\label{rf}  
\langle r\rangle_f=\frac{R^2}{dT}f. 
\end{equation}
More complicated forms of $p({\bf r})$ do not lead to a simple
linear relation, and in many cases the relation between $\langle r\rangle_f$ 
and an arbitrary $f$ cannot be calculated exactly. 
However, for sufficiently small $f$, linear response theory provides a 
simple universal answer: 
By expanding the exponent in Eq.~(\ref{Z}) in powers of $f$, 
and by omitting powers higher than 2, we see that Eq.~(\ref{rf}) is valid 
for arbitrary spherically symmetric $p({\bf r})$, provided that $R^2$  
is the mean squared EED calculated at zero force. 
The force can be considered small when $\langle r\rangle_f\ll R$, i.e., for $f\ll T/R$.   
 
The probability density of the EED of a phantom chain with a simple SL 
(Fig.~\ref{link_def}) is given by  
\begin{equation}\label{pr} 
p({\bf r})=p_{N-n}({\bf r})g_N(n) \, 
\end{equation} 
(see Eqs.~(\ref{pN}) and (\ref{gN})). Here, $N-n$ is the number of 
monomers in the {\em force-carrying backbone} of the polymer.  
Thus, $R^2$ can be found by integrating $r^2$ with the above statistical 
weight, over all possible ${\bf r}$ and $n$, leading to 
\begin{subequations}\label{Ree2} 
\begin{equation} 
R^2=a^2\left(N-\langle n\rangle_0\right),\label{Ree2a} 
\end{equation} 
where 
\begin{equation} 
\label{Ree2b} 
\langle n\rangle_0=  
\begin{cases} 
c_d & \text{for $d>4$}\\ 
c_4\ln N & \text{for $d=4$}\\ 
c_dN^{2-d/2} & \text{for $2<d<4$}\\ 
\frac{1}{2}\frac{N}{\ln N} & \text{for $d=2$}\\ 
\frac{2-d}{6-d}N & \text{for $d<2$} 
\end{cases} 
\qquad . 
\end{equation} 
\end{subequations} 
Here, $c_d$ are short length-scale cutoff-dependent constants. 
In this expression, $\langle n\rangle_0$ is simply the mean number of monomers inside 
the constricted loop, and $R^2$ is obtained by replacing $N$ in the expression 
for $R^2$ of a simple phantom chain by $N-\langle n\rangle_0$. 
We note that for $d>2$, $\langle n\rangle_0$ has a sub-linear dependence on $N$, 
and as $d$ increases the correction created by the SL 
depends more weakly on $N$ \cite{ba}. 
For $d>4$ random walks do  
not form long loops and $\langle n\rangle_0$ becomes independent of $N$.  
 
The expression for $R^2$ of a phantom chain with a SL 
can now be substituted in Eq.~(\ref{rf}), to obtain the FE relation 
$\langle r\rangle_f=f(a^2/dT)(N-\langle n\rangle_0)$ for small $f$. 
For large $f$, this expression is no longer valid. 
However, by direct inspection of the required average we find 
that Eqs.~(\ref{Z}), (\ref{rfZ}), and (\ref{pr}) lead to 
\begin{eqnarray}\label{rf_full} 
\langle r\rangle_f  
 &=& T\frac{\partial}{\partial f} \ln  
  \int dn\, g_N(n)\, \exp\left[\frac{f^2a^2(N-n)}{2dT^2}\right]  \nonumber \\ 
 &=&\frac{a^2}{dT}\left(N-\langle n\rangle_f \right) f\ , 
\end{eqnarray} 
where $\langle n\rangle_f$ is the mean loop size in the presence of the 
force, equal to 
\begin{equation}\label{nf_phantom} 
\langle n\rangle_f=\frac{\int_{n_0}^N dn\,n\, g_N(n)\exp\left[-f^2a^2n/(2dT^2)\right]} 
{\int_{n_0}^N dn\, g_N(n)\exp\left[-f^2a^2n/(2dT^2)\right]}\ . 
\end{equation} 
The lower limit of ${n_0}$ in the above integrals is the minimal loop size allowed by the 
specific model.

%%%%%%%%%%%%%%%%%%%%%%%%%%%%%%%%%%%%%%%%%%%%%%%%%%%%%%%%%%%%%%%%%%%%% 
\begin{figure} 
\includegraphics[width=8cm]{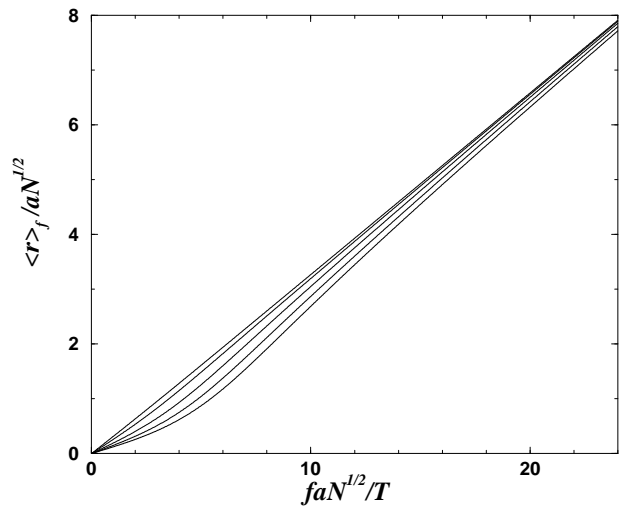} 
\caption{\label{rvsf_curv} 
Force-extension curves for a 100-monomer polymer with a  slip-link 
in $d=3$  with its loop threaded through additional (from top to bottom) 
$m=0$, 1, 2, 3 and 4,  rings.} 
\end{figure}  
%%%%%%%%%%%%%%%%%%%%%%%%%%%%%%%%%%%%%%%%%%%%%%%%%%%%%%%%%%%%%%%%%%%%% 

\begin{figure} 
\includegraphics[width=4cm]{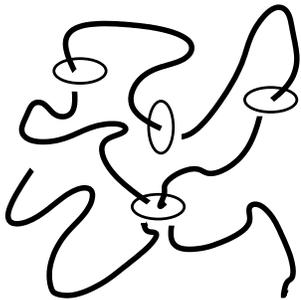} 
\caption{\label{many_links} A slip-link whose loop is threaded 
through an additional $m=3$ rings which are restricted to stay on  
the loop.} 
\end{figure}

The FE relation is no longer linear, 
even for a phantom polymer with one SL, although the deviation  
disappears rapidly with increasing $f$. 
For the simple SL with the weight in Eq.~(\ref{gN}), 
the nonlinearity is barely detectable even for the relatively small 
value of $N=100$, as indicated by the top line in Fig.~\ref{rvsf_curv}. 
In more complicated topologies of many SLs, we frequently encounter 
the behavior $g_N(n)\sim n^{-\alpha}$ \cite{metz}, where $n$ is the total
number of monomers that do not belong to the direct path between the ends
of the polymer, while $N$ appears in the prefactor or other non-singular
parts of the probability density. 
In such cases, the non-linearities become more pronounced as $\alpha$
decreases. Let us specifically consider the ``toy-example'' of a SL in
which $m$ additional rings 
slide around the loop, as depicted in Fig.~\ref{many_links}.
The number of ways of placing these sliding rings leads to
$g_N(N)={\cal N}(N-n)n^{m-d/2}$ \cite{remsd}. 
Depending on the values of $m$ and $d$, three different behaviors can be
distinguished: \par\noindent {\em (i)} For $m>d/2-1$, the integrals in
Eq.~(\ref{nf_phantom}) are dominated by large $n$. We can thus set the
lower limit of the integrals to 0, and introduce the new variable 
$x=n/N$, to get 
\begin{equation}\label{reduced} 
\frac{\langle r\rangle_f}{a\sqrt{N}}= 
\frac{fa\sqrt{N}}{T\sqrt{2d}}\left[ 
1-\frac{\int_0^1 dx\, x^{m+1-d/2}{\rm e}^{-xf^2a^2N/(2dT^2)}} 
{\int_0^1 dx\,x^{m-d/2}{\rm e}^{-xf^2a^2N/(2dT^2)}} 
\right]. 
\end{equation} 
This FE curve now satisfies a scaling form $\langle r\rangle_f/R=\Phi\left(
fR/[\sqrt{2d}T]\right)$, where the scaling function has the limits
\begin{equation}
\label{limits}
\Phi(z)\sim\begin{cases}
z/(1+m-d/2), & z\to0,\\
z-c/z, & z\to\infty.
\end{cases}
\end{equation}
The initial slope is reduced for larger $m$, as depicted in Fig.~\ref{rvsf_curv}, while
the asymptotic form at large $f$ is reached with a correction that falls off as $1/f$.
The physical origin of the nonlinearity is the tightening of the initially large loop
in the intermediate regime.

\par\noindent{\em (ii)}
For $d/2-2>m>d/2-1$, the mean loop size grows as $N^{2+m-d/2}$, playing
the role of an additional (sub-leading) length scale. Consequently, the FE curve
no longer has a simple scaling form, and behaves as
\begin{equation}
\label{ knot}
\frac{\langle r\rangle_f}{R}=\frac{fR}{\sqrt{2d}T}\left[1-N^{2+m-d/2}\phi\left(\frac{fR}{T}\right)\right].
\end{equation} 
This case may be most similar to that of knots in three dimensions, and indeed such 
corrections to scaling were used in Ref.~\cite{farago} to extract the size of the knot.
\par\noindent{\em (iii)}
For $m<d/2-2$, both integrals in Eq.~(\ref{nf_phantom}) are dominated by the
short distance cutoff, resulting in $\langle n\rangle_f\approx n_0$, independent of $f$.
The FE curve is thus linear in this regime, with finite size corrections that disappear as $1/N$.

\begin{figure} 
\includegraphics[width=8cm]{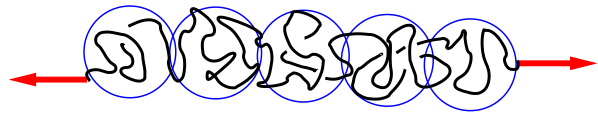} 
\caption{\label{blob} In the Pincus-de Gennes  scenario, the stretched 
polymer is viewed as a linear sequence of ``blobs'' (circumscribed by circles). 
Within each blob the polymer is unstressed; the size and number of  blobs 
depends on the stretching force.} 
\end{figure}

While the above results are easily obtained for a phantom 
polymer with a SL, it is convenient to restate them 
in a form that is more generally valid, and, in particular, 
applicable to interacting polymers. This will be done using 
the  Pincus-de Gennes picture \cite{blob}, 
according to which a stretched polymer (without a SL)
at short scales does not feel the influence of the external force,
and correlations remain as in the unforced polymer, 
while at longer distances it is essentially a linear object aligned to the force. 
The polymer can then be visualized as a linear chain of {\em blobs}, 
as depicted in Fig.~\ref{blob}. 
The number of monomers $N_b$  inside a blob is determined by the
condition $fR_b\approx T$, where $R_b$ is the EED of $N_b$ monomers. 
In the case of a phantom polymer ($R_b=aN_b^{1/2}$), 
this leads to $N_b\approx (T/fa)^2$, while in the more general case 
with $R_b=aN_b^\nu$, we get $N_b\approx (T/fa)^{1/\nu}$. 
Consequently, for large forces the EED of a whole polymer 
is the size of a single blob times the number of blobs, i.e., 
\begin{equation}\label{rfstrong} 
\langle r\rangle_f=(N/N_b)R_b=aNN_b^{\nu-1} 
=aN(fa/T)^{\frac{1}{\nu}-1}. 
\end{equation} 
For phantom polymers  the FE curve 
remains linear even for large $f$, while for a self-avoiding polymer in $d=3$ 
(with $\nu\approx0.58$) the relation is highly non-linear. 
 
%%%%%%%%%%%%%%%%%%%%%%%%%%%%%%%%%%%%%%%%%%%%%%%%%%%%%%%%%%%%%%%%%%%%% 
%%%%%%%%%%%%%%%%%%%%%%%%%%%%%%%%%%%%%%%%%%%%%%%%%%%%%%%%%%%%%%%%%%%%% 
 
\begin{figure} 
\includegraphics[width=8cm]{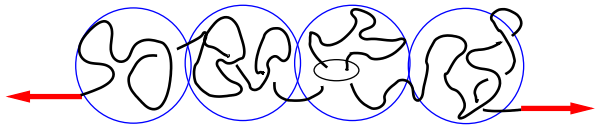} 
\caption{\label{link_blob} Qualitative representation of a stretched 
polymer with a SL. The loop created by the SL is  
contained within a single blob, and 
its size is determined only by the number of the monomers in the blob.} 
\end{figure}

We now note that $N_b$ is the scale over which the exponential 
factor in Eq.~(\ref{nf_phantom}) decays, and consequently 
the mean size of the SL loop for $f\gg T/(a\sqrt{N})$ 
can be estimated as  
\begin{equation}\label{nf_Nb} 
\langle n\rangle_f\approx \frac{\int_{n_0}^{N_b} dn\, n \, g_N(n)} 
{\int_{n_0}^{N_b} dn\,g_N(n)}\ . 
\end{equation} 
However, this expression is exactly the size of the link in a polymer consisting 
of $N_b$ monomers {\em in the absence of an external force}, i.e. 
\begin{equation}\label{fbe} 
\langle n\rangle_{f,N}\approx \langle n\rangle_{0,N_b}. 
\end{equation} 
The first subscript in this equation denotes the size of the force, 
while the second index indicates the total number of monomers. 
We can, therefore, view the SL loop as being confined to a single blob. 
Since within a blob the external force is not felt, its size is 
determined by regarding the entire polymer length as $N_b$,
as depicted qualitatively in Fig.~\ref{link_blob}. 
 
%%%%%%%%%%%%%%%%%%%%%%%%%%%%%%%%%%%%%%%%%%%%%%%%%%%%%%%%%%%%%%%%%%%%% 
%%%%%%%%%%%%%%%%%%%%%%%%%%%%%%%%%%%%%%%%%%%%%%%%%%%%%%%%%%%%%%%%%%%%% 
 
While the leading term in the expression for the EED of a strongly stretched  
polymer with a SL will still have a form given by Eq.~(\ref{rfstrong}), 
subleading corrections depend on the influence that the 
presence of SL has on the EED of the unstretched polymer. 
For a phantom chain there is a clear separation between the segment 
that creates the link and the remainder of the chain, and, consequently, 
in the absence of an external force, the reduction in $R^2$ 
can be simply related to the reduction of $N$ by Eq.~(\ref{Ree2}).  
This does not have to be the case in the presence of interactions,  
and each case must be considered separately. In the presence of  
self-avoiding interactions (for $d<4$), it can be shown \cite{metz}, 
that for $n\ll N$ the statistical weight of an $n$-monomer loop 
is given by $g_N(n)\sim n^{-c}$, where $c>2$. Consequently,  
$\langle n\rangle_0$ is independent of $N$, and will cause 
no detectable modification in the FE curve of a self-avoiding polymer. 
This picture can be easily generalized to a sequence of (non-interpenetrating) SLs.
At zero force, the loops will compete for the available length, each
acquiring a fraction of the overall length, as described in Ref.~\cite{metz}.
At strong force, in the blob regime, each slip link is most likely confined
to its own blob. 
 
Viewing the stretched polymer as a sequence of blobs, with only 
individual blobs affected by the presence of the constraints such as SLs, 
creates a convenient framework for evaluation of  FE relations. 
This picture may possibly be extended to knotted polymers: 
If the size of each prime knot factor depends on the number of monomers 
$N$ as a power law $N^t$, then 
the application of a strong stretching force will confine the knot 
to a blob and will reduce its size to $N_b^t$. 
Such a scenario (at small forces) was explored in a recent Monte Carlo study
\cite{farago}. 
 
Acknowledgments: We thank A. Hanke and O. Farago for helpful discussions.  
This work was supported by US-Israel Binational Science 
Foundation (grant 1999-007), and  by the National Science Foundation 
(grants DMR-01-18213 and PHY99-07949). 
RM acknowledges financial support from the Deutsche Forschungsgemeinschaft.


\begin{thebibliography}{25} 
 
\bibitem{catenanes}H. L. Frisch, and E.  Wasserman, J. Am. Chem. Soc.  
    {\bf 83}, (1961); K. Mislow, {\it Introduction to Stereochemistry}, 
    Benjamin, New York (1965); G. Schill, {\it Catenanes,  
    Rotaxanes, and Knots}, Academic, New York (1971); 
    D. M. Walba, Tetrahedron {\bf 41}, 3161 (1985); 
   {\em Catenanes, Rotaxanes and Knots}, edited by J.-P. 
   Sauvage and C. Dietrich-Buchecker,VCH, Weilheim (1999). 
 
\bibitem{alberts} B. Alberts, K. Roberts, D. Bray, J. Lewis, M. Raff and J. D. 
Watson, {\em The molecular biology of the cell} (Garland, New York, 1994). 
 
\bibitem{ringclosure}V. V. Rybenkov N. R. Cozazarelli, A. V. Vologodskii, 
     Proc. Natl. Acad.Sci. USA 
    {\bf 90}, 5307 (1993); S. Y. Shaw and J. C. Wang, Science  
    {\bf 260}, 533 (1993). 
 
\bibitem{enzyme}W. R. Bauer, F. H. C. Crick, and J. H. White, 
    Sci. Am. {\bf 243}, 118 (1980); N. R. Cozzarelli, S. J. Spengler,  
    and A. Stasiak, Cell {\bf 42}, 325 (1985);   
    S. A. Wasserman and N. R. Cozzarelli, Science {\bf 232}, 951 (1986). 
 
\bibitem{wang} H. Wang R. J. Di Gate and N. C. Seeman, 
Proc. Nat. Acad. Sci. (USA) {\bf 93}, 9477 (1996). 
 
\bibitem{taylor} W. R. Taylor, Nature {\bf 406}, 916 (2000); 
R. Takusagawa and K. Kamitori, 
J. Am. Chem. Soc. {\bf 118}, 8945 (1996). 
 
\bibitem{arai} T. Arai, R. Yasuda, K.-i. Akashi, Y. Harada, H. Miyata, 
K. Kinosita, Jr., and H. Itoh, Nature {\bf 399}, 446 (1999).   
 
\bibitem{mechsing} M. Rief, M. Gautel, F. Oesterhelt, J. M. Fernandez and 
H. E. Gaub, Science {\bf 276}, 1109 (1997);  
S. B. Smith, L. Finzi, and C. Bustamante, 
Science {\bf 258}, 1122 (1992); 
B. Smith, Y. Cui, and C. Bustamante, Science {\bf 271}, 795 (1996); 
K. Svoboda, S. M. Block, 
Ann. Rev. Biophys. Biomol. Structure {\bf 23}, 247 (1994); 
A. Ashkin, Proc. Natl. Acad. Sci. USA {\bf 94}, 4853 (1997);      
H. G. Hansma, J. Vac. Sci. Technology {\bf B14}, 1390(1995);
K. Berg-Sorensen and H. Flyvbjerg (unpublished).
 
\bibitem{degennesSC}P. G. de Gennes, {\it Scaling Concepts in Polymers 
Physics} (Cornell University Press, Ithaca, New York, 1979) 
 
\bibitem{edwards} M. Doi and S. F. Edwards, {\it The Theory of Polymer 
dynamics} (Clarendon Press, Oxford, 1986). 
 
\bibitem{mathdet}J. W. Alexander, Trans. Amer. Math.Soc. {\bf 30}, 275 (1928). 
V. F. R. Jones, Bull. Am. Math. Soc. {\bf 12}, 103 (1985);  
and Pacific J. Math. {\bf 137}, 311 (1989). 
P. Freyd, D. Yetter, J. Hoste, W. B. R. Lickorish, 
K. C. Millett and A. Ocneanu, Bull. am. Math. Soc. {\bf 12}, 103 (1985). 
L. H. Kauffman, Topology {\bf 26}, 395 (1987). 
 
\bibitem{slip}M. Doi and S. F. Edwards, J. Chem. Soc. Faraday Trans. 2 
    {\bf 74}, 1802 (1978);  
R. C. Ball, M. Doi, S. F. Edwards, and M. Warner, Polymer {\bf 22}, 1010 (1981); 
D. Adolf, Macromolecules {\bf 21}, 228 (1988); 
M. Kosc, Colloid Polym. Sci. {\bf 266}, 105 (1988); 
P. G. Higgs and R. C. Ball, Europhys. Lett. {\bf 8}, 357 (1989). 
 
\bibitem{metz} R. Metzler, A. Hanke, P. G. Dommersnes, Y. Kantor, M. Kardar,  
Phys. Rev. Lett. {\bf 88}, 188101 (2002); Phys. Rev. E (2002),
in press.
 
\bibitem{blob}P. Pincus, Macromol. {\bf 9}, 386 (1976); 
P. G. de Gennes, P. Pincus, R. M. Velasco, F. Brochard,  
J. Phys. (Paris), {\bf 37} 1461 (1976). 
 
\bibitem{d2}For $d=2$, ${\cal N}=1/N\ln N$, while for $d<2$,  
${\cal N}=[(2-d)(4-d)/4]N^{(d/2)-2}$.  
 
\bibitem{ba} The numerical prefactor $c_d$ will be  
of order unity when the diameter $b$ of the SL is of  
the same order of magnitude as the elementary step size $a$. 
For $b>a$, the prefactor will be of the order $(b/a)^{d-2}$, 
for $2<d<4$.

\bibitem{remsd}The short-distance scaling behavior of this example is in 
fact the same as with $m$ additional slip links sliding on the central 
loop \cite{metz}. 

\bibitem{farago}O. Farago, Y. Kantor, and M. Kardar, preprint (2002). 
 
\end{thebibliography}
\end{document}